\documentclass[preprint,proceedings]{rmaa}

\usepackage{paralist}

\usepackage{psfrag,color}

\SetYear{2005}
\SetConfTitle{11th Latin-American Regional IAU Meeting}

\title{A Dimensional Study of Disk Galaxies \\
(An excuse to Talk about the Hubble Sequence)} 

\author{
  X. Hernandez,\altaffilmark{1} 
  B. Cervantes-Sodi \altaffilmark{1}}

\altaffiltext{1}{Instituto de Astronom\'\i{}a, UNAM, C.U., M\'exico.} 

\shortauthor{X. Hernandez, B. Cervantes-Sodi}
\shorttitle{A Dimensional Study of Disk Galaxies}

\fulladdresses{

\item X. Hernandez: Instituto de Astronom\'\i{}a, Universidad Nacional
  Aut\'onoma de M\'exico, A. P. 70-264, 04510 Mexico, D.F.
  (\email{xavier@astroscu.unam.mx}).
}

\listofauthors{X. Hernandez, B. Cervantes-Sodi}
\indexauthor{Hernandez, X.}
\indexauthor{Cervantes-Sodi, B.}

\abstract{
We present a highly simplified model of the dynamical structure of a disk galaxy where only
two parmeters fully determine the solution, mass and angular momentum. We show through
simple physical scalings that once the mass has been fixed, the angular momentum parameter
$\lambda$ is expected to regulate such critical galactic disk properties as
colour, thickness of the disk and disk to bulge ratio. It is hence expected to
be the determinant physical ingeredient resulting in a given Hubble type. A simple
analytic estimate of $\lambda$ for an observed system is provided. An explicit comparison of
the distribution of several galactic parameters against both Hubble type and $\lambda$
is performed using observed galaxies. Both such distributions exhibit highly
similar characteristics for all galactic properties studied, suggesting $\lambda$
as a physically motivated classification parameter for disk galaxies.
}

\addkeyword{galaxies:fundamental parameters}
\addkeyword{galaxies:general}
\addkeyword{galaxies:structure}

\begin{document}
\maketitle

\section{Introduction}

The first thing an astronomer wants to know about a galaxy is its morphological type, typically
expressed through the Hubble classification scheme, introduced about 80 years ago.

The continued usage of the Hubble classification owes its success to the fact that on hearing 
the Hubble entry of a galaxy,
one immediately forms an image of the type of galaxy that is being talked about. A wide variety of physical
features of galaxies show good monotonic correlations with Hubble type, despite the presence of significant
overlap and dispersion. To mention only a few, total magnitudes decrease towards later types
 and colours become bluer while gas fractions diminish from early to late types (e.g. Roberts 
\& Haynes 1994). Bulge magnitudes and bulge to disk ratios decrease (e.g. Pahre et al. 2004) 
and disks become thinner (e.g. Kregel et al. 2005) 
in going to later types. 

However, galactic classification schemes suffer from several shortcomings. First, the 
diagnostics which enter into the assignment of type, do so in a fundamentally subjective manner. 
One has to take the images to
an expert who will somehow integrate various aspects of the galaxy in question to arrive at the type
parameter. That different authors generally agree, together with the correlations between type
and various galactic parameters is indicative of a solid physical substrate to galaxy classification
schemes, the particular nature of which however, has remained elusive. Second, type parameters
are essentially qualitative  
which make it difficult to relate type to definitive quantitative
aspects of a galactic system, and bring into question the validity of any statistical mathematical
analysis performed on galactic populations based on type. 
Finally, the type given to a galaxy is highly dependent on the information one has regarding the system.
This stems from the inputs that determine the galactic type, the relevance of the bulge, typically
redder than the disk, means that when observing at longer rest frame bands galaxies shift to 
earlier types, an effect compounded with the high wavelength sensitivity of the bright HII regions
which define the structure and morphology of the spiral arms. 
Similarly, projection effects become important, as the structure
of the arms completely disappears in edge-on disks, eliminating one of the diagnostics relevant to the
assignment of type.

Monotonic trends with Hubble type suggest that after having fixed the mass, there might exist one other
parameter whose variation gives rise to the Hubble sequence. Many approaches to galaxy
formation and evolution have appeared over the years, generally reaching the conclusion that it is 
the angular momentum of a galactic system what determines its main characteristics,
e.g. Fall \& Efstathiou (1980), Firmani et al. (1996).

\section{Theoretical Framework}
Having chosen the angular momentum as our second parameter we now construct a simple 
physical model for a galaxy which allows to estimate this parameter directly from observed galactic properties.

In the interest of capturing the most essential physics
of the problem, the model we shall develop will be the simplest one could possibly propose, little more than
a dimensional analysis of the problem.
We shall model only two galactic components, the first one a disk having a surface mass 
density $\Sigma(r)$ satisfying:

\begin{equation}
\label{Expprof}
\Sigma(r)=\Sigma_{0} e^{-r/R_{d}},
\end{equation} 

Where r is a radial coordinate and $\Sigma_{0}$ and $R_{d}$ are two constants which are allowed to vary from
galaxy to galaxy.
The second component will be a fixed dark matter halo having an isothermal $\rho(r)$ density 
profile, and responsible for establishing a rigorously flat rotation curve $V_{d}$ throughout 
the entire galaxy, such that

\begin{equation}
\label{RhoHalo}
\rho(r)={{1}\over{4 \pi G}}  \left( {{V_{d}}\over{r}} \right)^{2}, 
\end{equation}

and a halo mass profile $M(r)\propto r$.
We shall refer not to the total angular momentum $L$, but to the 
dimensionless angular momentum parameter

\begin{equation}
\label{Lamdef}
\lambda = \frac{L \mid E \mid^{1/2}}{G M^{5/2}},
\end{equation}

where $E$ is the total energy of the configuration. 
We now assume $\mid E \mid$ is given by 1/2 the potential energy of the halo,
which dominates through a halo to disk mass ratio $M_{H}/M_{d} = F \simeq 25$.
If the specific angular momenta of disk and halo are equal, $l_{d} =l_{H}$,
as would be the case for an initially well mixed protogalactic state, 
we can introduce a disk Tully-Fisher relation:
$M_{d}=A_{TF} V^{3.5}$ into to yield:

\begin{equation}
\label{LamObs}
\lambda=21.8 \frac{R_{d}/kpc}{(V_{d}/km s^{-1})^{3/2}}.
\end{equation}

Where $A_{TF}$ and $F$ were calibrated from Galactic parameters.
The above equation allows a direct estimate of
$\lambda$, a dimensionless numerical parameter with a clear physical interpretation for any observed galaxy, 
with no degree of subjectivity and little sensitivity to orientation effects.
 Also, given the Tully-Fisher relation, it is clear that
$V_{d}$ in eq(\ref{LamObs}) can be replaced for total magnitude on a given band, for cases where
the rotation velocity is not available.

We now study the expected scaling between $\lambda$ and other observables of a disk galaxy, 
for example colour, gas fraction, star formation activity, and degree of flatness of the disk. 
The main ingredient will be the Toomre parameter,

\begin{equation}
\label{Toomre}
Q(r)=\frac{\kappa(r) \sigma(r)}{\pi G \Sigma(r)},
\end{equation}

This parameter measures the degree of internal gravitational stability of the disk,
with $\kappa(r)$ accounting for the disruptive shears induced by the differential rotation,
$\sigma(r)$ modeling the thermal pressure of a component, both in competition with $\Sigma(r) G$, the self gravity
of the disk. Values of $Q \leq 1$ are interpreted as leading to gravitational instability in a 'cold'
disk. We replace $\kappa(r)$ in eq(\ref{Toomre}) with $\Omega(r) = V_{d}/r$, the angular velocity of the disk.        

The ratio of disk scale height to disk scale length, $h/R_{d}$, is a measurable characteristic 
of galaxies which it is easy to show, will scale with $\lambda$. Assuming a thin disk, virial equilibrium 
in the vertical direction yields a relation between $h$ and $\Sigma$,

\begin{equation}
\label{VertEq} 
h= \frac {\sigma_{g}^{2}} { 2 \pi G \Sigma}. 
\end{equation}

We use this relation for 
$h$ to replace the gas velocity dispersion appearing in equation (\ref{Toomre}) for a combination of 
$h$ and the surface density. The dependence on $\Sigma$ is replaced by a dependence on $M_{d}$, 
$M_{H}$ and $\lambda$ to get a new expression for the Toomre's stability criterion, which evaluating radial
dependences at $r=R_{d}$ yields,

\begin {equation}
Q^2= e 2^{5/2} \left( \frac{M}{M_{d}} \right) \left( \frac{h}{R_{d}} \right) \lambda
\end {equation}

With $F=1/25$, evaluating at $Q=1$, the stability threshold suggested by self-regulated star formation cycles, 
the ratio $h/R_{d}$ is obtained as:

\begin {equation}
\label{hRratio}
\frac{h}{R_{d}} =  \frac{1}{390 \lambda},
\end {equation}

\begin{figure}[!t]
  \includegraphics[width=\columnwidth]{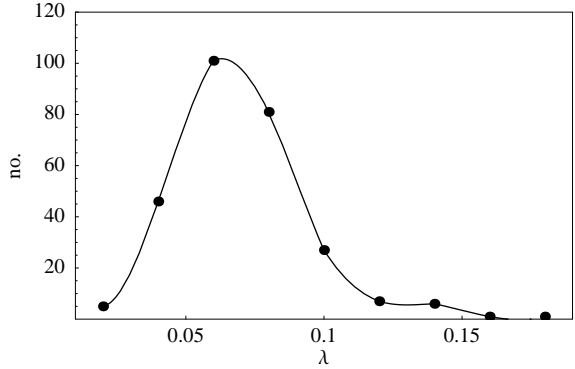}
  \caption{Distribution of values of $\lambda$ through eq(\ref{LamObs}) for the
complete ASSSG sample, with dots showing the binning.}
\end{figure}

Thinking of star formation as determined by a Schmidt law, $\dot{\Sigma}= \pi G \Sigma^2_{g}$,
we can use a characteristic time for the star formation process in the 
disk $\tau_{sf} = \Sigma_{g} / \dot{\Sigma}$, that gives an idea of the time a galaxy
can support a star formation rate of constant magnitude.
 From the above relation $\tau_{sf}$ becomes 

\begin{equation}
\label{tauSF}
\tau_{sf} = \frac{\pi G}{\Sigma_{g}}=\frac{2 \pi^2 G R^2_{d}}{M_{d}} \propto \lambda^2,
\end {equation}
 
at fixed disk mass. Alternatively, one could start from an
empirical Schmidt law of power $n$, a scaling $\tau_{sf} \propto \lambda^n$ will always result, which does not
alter our conclusions provided $n>1$, a general trait of Schmidt laws found in the literature.

Large values of $\lambda$ will correspond to long star formation periods; for this case we expect to have galaxies 
with young populations, looking relatively blue and having the high gas fractions of late type disks.
Galaxies with low $\lambda$ will have short star
formation periods; for them we will hardly see young stars and they will look red and gas poor, as
is the case for early types. A scatter
on this trend will be introduced by variations in total mass, together with a systematic reduction in
$\tau_{sf}$ in going to larger masses (c.f. equation \ref{tauSF}), in accordance with larger average
galactic masses found at earlier types.

\section{Data and Model Comparisons}

As mentioned in the preceding section, we expect a correlation between the $\lambda$ parameter and the 
Hubble type, since the analysis of section 2 leads one to suspect that it is this what 
primarily determines the properties of the galaxy
which are taken into account for its classification. 
We now test these expectations by direct comparison against two galaxy samples: the 
atlas of Courteau (1996) and the sample from Kregel et. al (2005), henceforth
ASSSG and dGKK, respectively.

In figure 1 we present the distribution of the 
ASSSG sample as a function of $\lambda$, estimated through eq(\ref{LamObs}). 
For the entire sample, the mean value is $< \lambda >=0.0645$. 
In general, the bulk of a galactic population moves to larger values of $\lambda$ for later types,
the mean value for each group is: $<\lambda>_{Sb}=0.0574, <\lambda>_{Sbc}=0.0649, <\lambda>_{Sc}=0.0689$.

We see in figure 1 an approximately Gaussian shape, with a considerable extension towards the larger values of
$\lambda$. We do not explore the details of this distribution further, as the ASSSG sample is not
intended to be complete or statistically unbiased in any sense. However, use of eq(\ref{LamObs})
in one such sample should prove useful in obtaining an empirical distribution of $\lambda$ parameters,
to compare for example, against predictions from cosmological models.

\begin{figure}[!t]
  \includegraphics[width=\columnwidth]{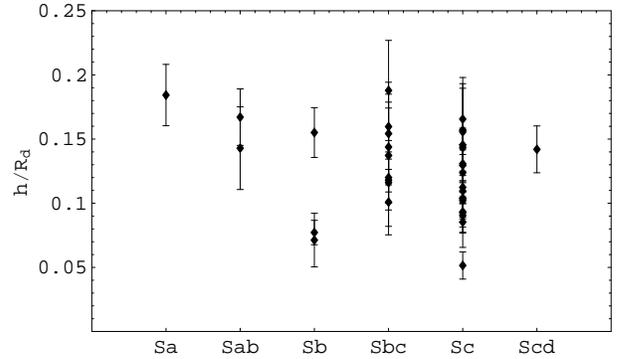}
  \caption{$h/R_{d}$ ratios for galaxies according to their Hubble type,
for galaxies in the dGKK sample.}
\end{figure}

The analysis of the distribution of $h/R_{d}$ ratios was made with the dGKK sample. 
It contains galaxies with 
more morphological types; from Sa to Scd galaxies. Figure 2 shows early galaxies 
presenting varied values 
for the $h/R_{d}$ ratio, although the means show a clear diminishing trend in going to latter types. 
Late type galaxies at the Sc end clearly show
on average much thinner disks than the earlier types, as confirmed by several authors over the last few years.

The complementary plot to figure 2 is given in figure 3, where instead of presenting 
$h/R_{d}$ vs $\lambda$, we use a logarithmic plot to better detect the presence of a relation
between these two parameters of the type the dimensional analysis of section 2 leads us to expect, 
equation(\ref{hRratio}). The line drawn in figure 3 is the best fit straight line for the sample,
having a slope of $-0.4 \pm 0.9$, not at odds with value of $-1$ predicted by the simple model. 
Trends for colour distributions and bulge to disk ratios (not shown) are also qualitatively
very similar when plotting against Hubble type or $\lambda$, Hernandez \& Cervantes-Sodi (2006).

We must now check our estimate of $\lambda$ in cases where this parameter is in fact known, galaxy formation
simulations where $\lambda$ acts as an input parameter. We use results from two groups, 
models calculated in connection with a study of star formation in disk galaxies applied to the 
Milky Way (Hernandez et al. 2001), and the models published by van den Bosch (2000).

\begin{figure}[!t]
  \includegraphics[width=\columnwidth]{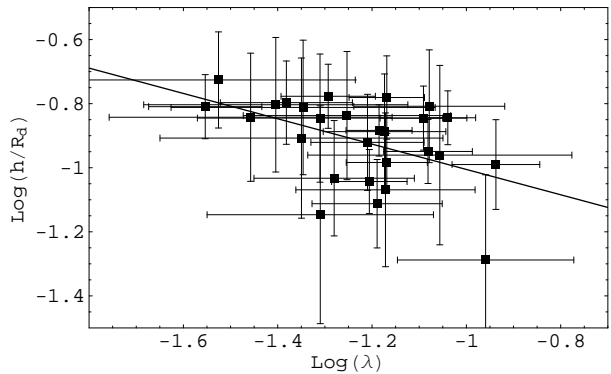}
  \caption{Logarithmic plot of the relation between $\lambda$ and $h/R_{d}$,
for galaxies in the dGKK sample, the solid line shows a linear fit to the data, yielding
a slope of $-0.4 \pm 0.9$.}
\end{figure}

Both models simulate the cosmological dark matter and baryonic mass accretion through a primordial 
fluctuation spectrum, and the subsequent, self-consistent dynamical and star formation evolution of a galaxy. 
The details of both models
vary in numerical approaches, resolution, time step issues and the level of approximation and inclusion of the
many different physical aspects of the complicated problem. Figure 4 shows the $\lambda_{model}$
values given as input by the above authors to their detailed formation codes, with the y-axis showing
$\lambda$ from equation (\ref{LamObs}) 
applied to the final results of the code evolution which followed
the model for $\sim 13$ Gyr.


It is interesting that the particular features of a galactic formation scenario
which we are interested in treating here, $\lambda$, final resulting $R_{d}$ and flat rotation region $V_{d}$,
are evidently very well modeled by the trivial physics that went into eq(\ref{LamObs}), across the more than
2.5 orders of magnitude covered by the masses in the modeled galaxies. 
This can be understood by considering that eq(\ref{LamObs}) is the result of two fundamental hypothesis: 
\adjustfinalcols
I) an exponentially decreasing surface
density profile for the disk, and II) a dominant dark halo responsible for a flat rotation curve,
linked to the disk through a Tully-Fisher relation. 
Whenever the above two conditions are met, as is the case for the final state of all published modeled galaxies and
real observed systems, basic physics strongly constrains results to lie not far from equation (\ref{LamObs}).

\begin{figure}[!t]
  \includegraphics[width=\columnwidth]{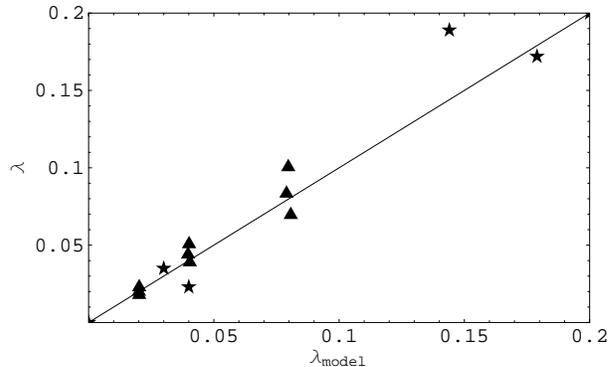}
  \caption{A comparison of the input value of $\lambda$ in a series of cosmological
CDM galactic evolution scenarios, $\lambda_{model}$, and $\lambda$ calculated on the final result of the
given simulations through eq(\ref{LamObs}). The triangles are models from Hernandez et al. (2001) and stars from 
van den Bosh (2000).}
\end{figure}

To conclude, we have presented an easily applicable way of estimating the $\lambda$ parameter, a theorist's
favourite descriptor of a galaxy type, for any real observed disk galaxy, which is therefore proposed as a 
physically motivated, objective classification parameter for disk galaxies. An extensive development of these
ideas can be found in Hernandez \& Cervantes-Sodi (2006).

\end{document}